\documentclass[aps,prl,preprint,groupedaddress]{revtex4-1}
\usepackage{amsmath}

\usepackage{color}

\def\al{\alpha}
\def\be{\beta}
\def\ga{\gamma}
\def\de{\delta}
\def\ep{\epsilon}

\def\la{\lambda}

\def\si{\sigma}

\def\La{\Lambda}

\def\mn{{\mu\nu}}

\def\half{{\textstyle{1\over 2}}}
\def\quar{{\textstyle\frac14}}
\def\frac#1#2{{\textstyle{{#1}\over {#2}}}}

\def\lsim{\mathrel{\rlap{\lower4pt\hbox{\hskip1pt$\sim$}}
    \raise1pt\hbox{$<$}}}
\def\gsim{\mathrel{\rlap{\lower4pt\hbox{\hskip1pt$\sim$}}
    \raise1pt\hbox{$>$}}}
\def\sqr#1#2{{\vcenter{\vbox{\hrule height.#2pt
         \hbox{\vrule width.#2pt height#1pt \kern#1pt
         \vrule width.#2pt}
         \hrule height.#2pt}}}}

\newcommand{\beq}{\begin{equation}}
\newcommand{\eeq}{\end{equation}}
\newcommand{\bea}{\begin{eqnarray}}
\newcommand{\eea}{\end{eqnarray}}

\begin{document}

%\preprint{}

\title{Gupta-Bleuler Photon Quantization in the SME}


\author{Don Colladay and Patrick McDonald}
%\thanks{}
\affiliation{New College of Florida, Sarasota, FL}
\author{Robertus Potting}
\affiliation{CENTRA and Department of Physics, University of the Algarve,
8005-139, Faro, Portugal}


\begin{abstract}
Photon quantization is implemented in the standard model extension (SME)
using the Gupta-Bleuler method and BRST concepts.  The quantization prescription applies
to both the birefringent and non-birefringent CPT-even couplings.  
A curious incompatibility is found between the presence of the Lorentz-violating terms
and the existence of a nontrivial conjugate momentum $\Pi^0$ yielding problems with 
covariant quantization procedure.  
Introduction of a mass regulator term can avoid the vanishing of 
$\Pi^0$ and allows for the implementation of a covariant quantization procedure.
Field-theoretic calculations involving the SME photons can then be performed using the mass
regulator, similar to the conventional procedure used in  electrodynamics
for  infrared-divergence regulation.
\end{abstract}

\pacs{}
%\keywords{}

\maketitle


\section{Introduction}
The Standard Model Extension (SME) is a framework for incorporating Lorentz and
CPT-violating effects into the Standard Model \cite{colkos1} \cite{colkos2}.  
While issues of quantization in the fermion sector have received significant attention
\cite{fermion-quant}, 
the corresponding problem in the photon sector has not been sufficiently addressed
due to a variety of difficulties that arise involving compatibility of gauge invariance,
Lorentz violation, and non-orthogonality of the polarization vectors.
Some recent progress has been made,  involving a perturbative approach to the
non-birefringent case \cite{hpw}, but the
birefringent case remains largely unaddressed. 
A recent study on the subject involves a one-parameter model
\cite{shreck}. 
It is the goal of this paper to help fill in this gap in the literature and outline the procedure for
quantizing the photon sector of the SME in the presence of CPT-even terms.  The CPT-odd 
terms can be problematic even at tree level, so their quantization is not discussed in this
paper.

Stringent bounds have been placed on Lorentz and CPT-violating parameters in the photon
sector \cite{kosrussel} with bounds on birefringent terms significantly stronger than non-birefringent
terms, mainly due to the possibility of astrophysical tests for the former.
Frequently birefringent terms are neglected in theoretical treatments due to this fact.
I spite of this, the birefringent terms remain an interesting theoretical tool
for studies of symmetry violation and present interesting challenges.
Additional motivation arises from the fact that birefringence phenomena are in fact observed
in various crystal substances.

Attempting to directly quantize
the SME photon with zero mass runs into some
serious impediments involving a lack of completeness of the polarization states, even when a 
gauge-fixing term is included in the lagrangian.
One implication of this can be that the conjugate momentum $\Pi^0$ introduced by the gauge-fixing
term can be driven identically to zero by the perturbed equations of motion.
It turns out that by adding a photon mass, these issues can be alleviated and the covariant
quantization procedure may be carried out consistently.
For this reason, a photon mass is included in our calculations.  
Field-theoretic computations involving the photon should then be possible where the zero mass limit is taken
at the end of any calculation, similar to what happens in the infrared divergence renormalization 
procedure in standard QED.

The paper is organized as follows.
Section two presents the photon sector of the SME and
analyzes properties of the solutions of the equation of motion.
Section three
gives the quantization rules for the theory.  
A central result involving the modified orthogonality properties of the polarization vectors
is used to invert the Fourier transform of the fields leading to a conventional algebra of raising
and lowering operators that create and annihilate photon modes in the vacuum.
Section four presents a computation of the propagator and includes a verification that the
conventional Green's function method applies in this context.
Section five discusses
the Hamiltonian formalism and verifies the correct action of the energy and momentum 
operators on the state space.
Section six presents a classification of $k_F$ terms according to their duality properties.
Finally, an explicitly solvable one-parameter example is presented to illustrate the formalism.  
The appendix includes an analysis of BRST symmetry to the theory involving an 
additional Stueckelberg field that can be used to find the physical subspace of the larger Hilbert
space with indefinite metric.

\section{Photon sector of the SME}
The photon sector of the SME consists of a CPT-even tensor $(k_F)^{\mu\nu\al\be}$
and a CPT-odd vector $(k_{AF})^\mu$.
Only the CPT-even tensor will be considered here as the CPT-odd
tensor can lead to negative energy problems even in the tree-level theory \cite{jackiw}.
Covariant gauge fixing in the photon sector can be implemented
in the usual way leading to the lagrangian
\cite{cambiaso}
\beq
\mathcal{L}_A = -{1 \over 4}F_\mn F^{\mn} - {1 \over 4} k_F^{\mu \nu \al \be} F_{\mn} F_{\al \be} 
+{1 \over 2} m^2 A_\mu A^\mu - {1 \over 2 \xi}(\partial_\mu A^\mu)^2 .
\label{lagrangian-A}
\eeq
The choice $\xi=1$ (Feynman gauge) is particularly convenient for the Gupta-Bleuler
quantization procedure \cite{gupta,bleuler} and is used in the rest of the paper.
Using the Stueckelberg method, a mass term for the photon
has also been included in (\ref{lagrangian-A})
to avoid certain issues that can arise
in the quantization of the massless theory, as we will see below.
The equation of motion in momentum space is
\beq
\left[ (p^2 - m^2)\eta^\mn  + 2 (k_F)^{\mu \al \nu \be} p_\al p_\be \right] \ep_\nu = 0,
\label{eom}
\eeq
where $\ep^\mu$ is the polarization vector.
One implication of this equation is found by dotting with $p_\mu$,
yielding the condition 
\beq
(p^2-m^2) (\ep \cdot p) = 0 ,
\label{transversality}
\eeq
indicating that states that satisfy a perturbed dispersion relation must be transverse
and
any state with $\ep \cdot p \ne 0$
must satisfy the conventional
Lorentz covariant dispersion relation.
Setting the determinant of the coefficient matrix
of (\ref{eom}) to zero
yields an eighth-order polynomial in $p^0$ that must
be solved to yield the energies.  
This dispersion relation can be expressed in terms of scalar invariant powers of
the perturbation $K^{\mn} = 2 k_F^{\mu\al\nu\be}p_\al p_\be$.
For example, when $K^\mu{}_\mu = 0$ 
(this is the case for $k_F$ anti-self-dual and therefore birefringent, 
a fact to be explained later in the paper),
the dispersion relation takes the relatively simple form
\newcommand{\Tr}{\mbox{Tr}\,}
\beq
\left[ p^2 - m^2 \right] \left[ 6(p^2 - m^2)^3 - 3 (p^2 - m^2)\Tr K^2 + 2\Tr K^3 \right] = 0 ,
\eeq
where $\Tr K^2 = K^\mu{}_\nu K^\nu{}_\mu$ and 
$\Tr K^3 = K^\mu{}_\nu K^\nu{}_\al K^\al{}_\mu$.
Note that $K^\mn p_\nu = 0$ indicating that the fourth-order term must vanish as the
null-space of $K$ is always at least one dimensional, implying that its determinant is 
identically zero.
A symmetry under $p^\mu \rightarrow -p^\mu$
reduces the number of independent solutions to four after reinterpretation of 
the negative energy solutions as positive energy ones.
We also note that a factor of $p^2$ can be extracted from the $\Tr K^3$ term
using an argument discussed in \cite{kostmewes} involving the dual action of $\La^3 K$ on the momentum vector.
This means that in the massless limit,
a factor of $p^4$ can be factored out of the dispersion relation,
in agreement with the results of \cite{kostmewes}.
This equation (for $m \ne 0$) normally has four independent solutions which are labeled using
$\la=0,1,2,3$.  
An orthogonality relation can be derived using the equation of motion in the form
\beq
\left[ \omega_p^2\eta^\mn + 2 k_F^{\mu 0 i \nu}p_i p_j \right] \ep_\nu^{(\la)}(\vec p)
= \left[ (p_0^{(\la)})^2 \tilde \eta^\mn - 2 \hat k_F^{\mu 0 i \nu} p_0^{(\la)} p_i \right] \ep_\nu^{(\la)}(\vec p),
\eeq
with $\omega_p = \sqrt{\vec p^2 + m^2}$ and 
$\hat k_F^{\mu 0 i \nu} = k_F^{\mu 0 i \nu} + k_F^{\nu 0 i \mu}$.
Note that all of the dependency on $p_0$ has been moved to the right-hand side of the equation.
This equation can then by multiplied by $\ep^{(\la^\prime)}_\mu(\vec p)$ on the left
and subtracted from the corresponding equation with $\la \leftrightarrow \la^\prime$
switched.  The result is the following orthogonality relation which is valid
when $p_0^{(\la)} \ne p_0^{(\la^\prime)}$
\beq
\ep^{(\la^\prime)}_\mu(\vec p) \left[ \left( p_0^{(\la)} + p_0^{( \la^\prime)}\right) \tilde \eta^\mn 
-2 \left(k_F^{\mu 0 i \nu} + k_F^{\nu 0 i \mu} \right) p_i \right] \ep_\nu^{(\la)}(\vec p)
=2 p_0^{(\la)} \eta^{\la \la^\prime},
\label{orthog1}
\eeq
and the normalization has been chosen to match the conventional case
with $\la=0$ labeling the timelike vector.
A similar orthogonality relation exists between polarization vectors with opposite
three-momenta
\beq
\ep^{(\la^\prime)}_\mu(-\vec p) \left[ \left( p_0^{(\la^\prime)}(- \vec p) 
+ p_0^{( \la)}(\vec p)\right) \tilde \eta^\mn 
+2 \left(k_F^{\mu 0 i \nu} + k_F^{\nu 0 i \mu} \right) p_i \right] \ep_\nu^{(\la)}(\vec p)
=0.
\label{orthog2}
\eeq
Note that for parity-violating coefficients ($k_F^{0ijk}$)
one has generally
$p_0^{(\la)}(\vec p) \ne p_0^{(\la)}(-\vec p)$ in contrast to the conventional case,
so care must be taken to distinguish the various quantities.

\section{Quantization}
The canonical momentum is computed in the usual way by taking derivatives of
Lagrangian (\ref{lagrangian-A}) with respect to the $\dot A_\mu$ fields 
\beq
\Pi^j = F^{j0} + k_F^{j0 \alpha \beta}F_{\al \be}, \quad \Pi^0 = - \partial_\mu A^\mu .
\eeq
Imposing equal-time canonical commutation rules
\beq
[A_\mu(t,\vec x), \Pi^\nu(t,\vec y)] = i g_\mu^{~\nu} \de^3(\vec x - \vec y),
\label{comrels0}
\eeq
along with
\beq
[A_\mu(t,\vec x), A_\nu(t,\vec y)] = [\Pi^\mu(t,\vec x), \Pi^\nu(t,\vec y)]=0 ,
\label {comrels1}
\eeq
implements the standard canonical quantization in a covariant manner as is done in
the conventional Gupta-Bleuler method.
This implies that the time derivatives of the spatial components $A^i$ satisfy
the modified commutation relations
\beq
[\dot A^i(t,\vec x), A^j(t,\vec y)] = - i  R^{ij} \de^3(\vec x - \vec y)
\eeq
where $R^{ij}$ is the inverse matrix of $\de^{ij} - 2 (k_F)^{0i0j}$.
In any concordant frame where $k_F$ is reasonably small, this inverse exists.
The commutation relations involving $\dot A^0$ and $A^i$are the same as in the usual
case, so for future use, it is convenient to define a covariant-looking
tensor $\overline \eta^\mn$ by setting
$\overline \eta^{00} = 1$, $\overline\eta^{0i} = 0$, and $\overline \eta^{ij} = -R^{ij}$.
The commutation relations take the form
\beq
[\dot A^\mu(t,\vec x), A^\nu(t,\vec y)] = i \overline\eta^\mn \de^3(\vec x - \vec y) .
\label{comrels2}
\eeq
This matrix is also the inverse of $\tilde \eta^\mn = \eta^\mn -2 k_F^{\mu 0 0 \nu}$
as $\overline\eta^\mn \tilde\eta_{\nu \al} = \eta^\mu_{~\al}$.
Note also that the time derivatives of $A$ no longer commute, rather
\beq
[\dot A^\mu(t,\vec x), \dot A^\nu(t,\vec y)] = -2 i \overline \eta_{\mu \alpha}
(k_F^{\al 0 i \be} + k_F^{\be 0 i \al})\overline \eta_{\be \nu}{\partial \over \partial x^i}
\de^3(\vec x - \vec y),
\label{comrels3}
\eeq
involving the spatial derivatives of the delta function.

The field can then be expanded in terms of Fourier modes as
\beq
A_\mu(x) = \int {d^3 \vec p \over (2 \pi)^3} \sum_\la {1 \over 2 p_0^{(\la)}}\left( a^\la(\vec p) 
 \ep_\mu^{(\la)}(\vec p)
e^{-i p \cdot x} +  a^{\la\,\dagger} (\vec p)  \ep_{\mu}^{(\la)}(\vec p)
e^{i p \cdot x}\right).
\label{fouriermodes}
\eeq
This formula may be inverted using the orthogonality relations (\ref{orthog1}) and (\ref{orthog2}) as
\beq
a^{\la}(\vec q) = \eta^{\la \la^\prime} \int d^3 \vec x e^{i q\cdot x} \left[ i {\stackrel{\leftrightarrow}{\partial_0}}\tilde\eta^\mn
 - 2 (k_F^{\mu 0 i \nu} + k_F^{\nu 0 i \mu})q_i \right] \ep_\nu^{(\la^\prime)}(\vec q)A_\mu(x),
\eeq
and
\beq
 a^{\la\,\dagger}(\vec q) = \eta^{\la \la^\prime} \int d^3 \vec x\,
 e^{-i q\cdot x} \left[ -i {\stackrel{\leftrightarrow}{\partial_0}}\tilde\eta^\mn
 - 2 (k_F^{\mu 0 i \nu} + k_F^{\nu 0 i \mu})q_i \right] \ep_\nu^{(\la^\prime)}(\vec q)A_\mu(x).
\eeq
These can be used to compute
\beq
[a^\la(\vec p), a^{\la^\prime\,\dagger}(\vec q)] = - (2 \pi)^3 2 p_0^{(\la)} \eta^{\la \la^\prime} \de^3(\vec p - \vec q),
\label{a-adagger}
\eeq
as well as 
\beq
[a^\la(\vec p), a^{\la^\prime}(\vec q)] = 
[a^{\la\,\dagger}(\vec p), a^{\la^\prime\,\dagger}(\vec q)] = 0,
\label{a-a}
\eeq
demonstrating that the raising and lowering operators obey conventional bosonic statistical relations.

It remains to impose the restriction on the physical states that eliminates the negative norm states
as in the conventional Gupta-Bleuler approach.
As is shown in the appendix,
the physical states  still satisfy the usual condition

\begin{equation}
\langle \psi_{\rm phys}| \partial_\mu A^\mu |\psi_{\rm phys} \rangle = 0.
\end{equation}

Application of the basic commutators (\ref{a-adagger}) and (\ref{a-a})
to the coordinate-space commutation relations 
(\ref{comrels1}), (\ref{comrels2}), and (\ref{comrels3})
for the vector potential imply a variety of modified completeness relations 
for the polarization vectors, including
\beq
{1 \over 2}\sum_{\la,\la^\prime} \eta_{\la \la^\prime}\left[ \ep_\mu^{(\la)}(\vec p) \ep_\nu^{(\la^\prime)}(\vec p) 
+ \ep_\mu^{(\la)}(-\vec p) \ep_\nu^{(\la^\prime)}(-\vec p)  \right] = \overline \eta_{\mn}.
\label{completenessrel}
\eeq
Note that the right-hand side is independent of momentum indicating that a complete
set of polarization vectors must exist for every momentum choice.  
An explicit example is presented later in the paper for which the above completeness
relation holds.  
Unfortunately, this is not always true in the massless limit as will be seen in the explicit example, 
so the presence of a mass term is crucial for the consistency of the quantization procedure.

\section{The propagator}
In Lorentz-invariant field theory the vacuum expectation value
of the time-ordered product of two field operators plays a central role
as it is a Green's function (the Feynman propagator).
It is important to check that this is still the case
for the time-ordered product $\langle0|TA_\mu(x)A_\rho(y)|0\rangle$.
Indeed, by acting with the kinetic operator one obtains
\begin{align}
\bigl((\partial^2+m^2)\eta^{\mu\beta}-&2k_F^{\mu\nu\alpha\beta}\partial_\nu
\partial_\alpha\bigr)\langle0|A_\beta(x)A_\rho(y)\theta(x^0-y^0)
+A_\rho(y)A_\beta(x)\theta(y^0-x^0)|0\rangle=\nonumber\\
&=\delta(x^0-y^0)(\eta^{\mu\beta}\eta^{0\alpha}
-k_F^{\mu0\alpha\beta}-k_F^{\mu\alpha0\beta})
\langle0|[\partial_\alpha A_\beta(x),A_\rho(y)]|0\rangle\nonumber\\
&=\delta(x^0-y^0)\tilde\eta^{\mu\beta}
\langle0|[\dot A_\beta(x),A_\rho(y)]|0\rangle\nonumber\\
&=\delta(x^0-y^0)\tilde\eta^{\mu\beta}i\bar\eta_{\beta\rho}
\delta^3(\vec x-\vec y)\nonumber\\
&=i\delta^4(x-y)\delta^\mu_\rho,
\label{time-ordered-product}
\end{align}
verifying that in fact the time-ordered product still functions as the appropriate
Green's function for the perturbed equation of motion.
In the first identity we used the fact that $A_\mu(x)$ satisfies the
equation of motion, so that the only nonzero contributions
arise when at least one time-derivative acts on the delta-functions.

Using the mode expansion (\ref{fouriermodes}), it follows that
\begin{align}
\langle0|TA_\mu(x)A_\nu(y)|0\rangle=-\int{d^3p\over(2\pi)^3}
&\sum_{\lambda,\lambda'}{\eta_{\lambda\lambda'}\over 2p_0^{(\lambda)}(\vec p)}
\ep^{(\lambda)}_\mu(\vec p)\ep^{(\lambda')}_\nu(\vec p)\nonumber\\
&{}\times\left(e^{-ip\cdot(x-y)}\theta(x^0-y^0)+e^{ip\cdot(x-y)}\theta(y^0-x^0)
\right),
\label{propagator-mode-expansion}
\end{align}
expressing the propagator explicitly in terms of a polarization sum.
Eq.\ (\ref{propagator-mode-expansion}) complements the covariant
expressions obtained in \cite{cambiaso} for the propagator.

\section{Hamiltonian Formalism}

Calculation of the hamiltonian density
${\cal H} = \Pi^\mu \dot A_\mu - {\cal L}$
yields (after appropriate partial integrations with $H = \int d^3 \vec x {\cal{H}}$),
\bea
{\cal H} & = &{1 \over 2}\left( (\partial_j A^j)^2 - (\dot A^0)^2+A^0\partial_0 (\partial \cdot A) + \vec E^2 + \vec B^2 \right) \nonumber \\
& & -k_{0i0j}F^{0i}F^{0j} + {1 \over 4} k_{ijkl} F^{ij} F^{kl} - {1 \over 2}m^2 A_\mu A^\mu .
\eea
The commutation relations fix the action of the hamiltonian on fields to be
the conventional relation
\beq
i [H,A_\mu] = {\partial_0 A_\mu } ,
\label{hamact}
\eeq
ensuring that the appropriate zero components of momentum are the energy 
eigenvalues.
The momentum operator can be constructed using the conserved energy-momentum
tensor as
\beq
P^i = \int d^3 \vec x \left( \Pi^j \partial^i A_j - (\partial \cdot A)\partial^i A^0 \right),
\eeq
and satisfies
\beq
i[P^i, A_\mu] = {\partial^i A_\mu}.
\eeq

\section{Classification of $k_F$ terms}
A useful classification of the $k_F$ tensors involves a splitting of the tensor
into self-dual and anti-self dual pieces.  
\beq
k_F = k_F^{SD} \oplus k_F^{ASD} ,
\eeq
where the dual of $k_F$ is defined by
\beq
(\tilde k_F)^{\mn \al \be} = 
{1 \over 4} \ep^{\mn \rho \si} \ep^{\al \be \ga \de}(k_F)_{\rho \si \ga \de} .
\eeq
A straightforward calculation reveals that the self-dual condition 
$k_F = \tilde k_F$ corresponds exactly to the nonzero single trace components of $k_F$, while the
anti-self dual condition corresponds to the trace-free condition on $k_F$.
One other useful fact is that the $F^2$ term in the lagrangian corresponds to 
a choice of an anti-self dual $k$-term of the form
\beq
\eta^{\mn \al \be} \equiv {1 \over 2} \left( \eta^{\mu \al} \eta^{\nu \be} - \eta^{\mu \be} \eta^{\nu \al} \right),
\eeq
which is typically set to zero to avoid a Lorentz-preserving correction to the kinetic term
of the lagrangian.
The trace components can be most easily handled using coordinate redefinition
techniques \cite{colmcdonfredef}, at least when dealing with the free theory.  
Application of the anti-self dual condition on $k_F$ implies that the $(k_F)^{ijkl}$
coefficients can be expressed in terms of the $(k_F)^{0i0j}$.
In addition, the only nonvanishing $(k_F)^{0ijk} $ components are the ones 
with $i \ne j\ne k$.
This decomposition has previously been pointed out in
\cite{colmcdoninst} where the trace terms were anti-self dual, opposite the current application, due
to the fact that Euclidean space was used in the instanton theory rather than
Minkowski space.

\section{Illustrative Single-Parameter Example}
The above formalism is well-illustrated by the following exactly solvable example involving only one
non-vanishing parameter.
The fundamental impediment to covariant quantization of the massless photon
can also easily be seen in this example.  
The $k_F^{\mn\al\be}$ tensor is selected
so that $k_F^{0101} = k/2$, and $k_F^{2323} = - k/2$ to satisfy the anti-self-dual condition.
The natural symmetries of $k_F$ are also imposed yielding $k_F^{1001} = - k_F^{0101} = \cdots$.
The double trace term $(k_F)^\mn_{~~\mn}$ is not zero for this choice, but it can be made zero 
by adding an appropriate overall trace term to $k_F$
\beq
(k_F^\prime)^{\mn\al\be} = (k_F)^{\mn \al \be} +{k \over 3} \eta^{\mn\al\be},
\eeq
where $\eta^{\mn \al \be} = (1/2) \left( \eta^{\mu \al} \eta^{\nu \be} - \eta^{\mu \be} \eta^{\nu \al} \right)$.

The dispersion relation factors nicely into four separate terms
\beq
(p^2 - m^2)(p_0^2 - \omega_{1})(p_0^2 -  \omega_{2}^2)(p_0^2 - \omega_{3}^2) = 0,
\eeq
where $ \omega_{1}^2 = \vec p^2 + m^2/(1 + k/3)$, 
$\omega_{2}^2 = \vec p^2 + (m^2 + k (p_2^2 + p_3^2))/(1-2k/3)$,
and $ \omega_{3}^2 = \vec p^2 + (m^2 - k(p_2^2+ p_3^2))/(1 + k/3) $,
are perturbed versions of $\omega_p^2 = \vec p^2 + m^2$.
The polarization vectors associated with each energy are,
\beq
\ep^{(0)}_\mu = {1 \over m}
\left(
\begin{array}{c}
\omega_p \\
p_1\\p_2\\p_3
\end{array}
\right),
\quad
\ep^{(1)}_\mu = {1 \over m \sqrt{(\omega_1^2 - p_1^2) (p_2^2 + p_3^2)}}
\left(
\begin{array}{c}
\omega_{1}(p_2^2 + p_3^2)\\ p_1 (p_2^2 + p_3^2)\\p_2( \omega_{1}^2 - p_1^2) 
\\ p_3( \omega_{1}^2 - p_1^2) 
\end{array}
\right),
\eeq
\beq
\ep^{(2)}_\mu = {{1 \over \sqrt{(1 - 2k/3)( \omega_{2}^2 - p_1^2})}}
\left(
\begin{array}{c}
p_1\\  \omega_{2} \\ 0 \\ 0
\end{array}
\right),
\quad
\ep^{(3)}_\mu = {1 \over \sqrt{(1 + k/3)(p_2^2 + p_3^2)}}
\left(
\begin{array}{c}
0\\0\\- p_3 \\ p_2
\end{array}
\right),
\eeq
where the normalizations
are chosen so that Eq.\ (\ref{completenessrel}) is satisfied.
When the mass is nonzero, there are always one timelike and three spacelike polarization
vectors.  
When the mass is identically zero, an issue can be seen by examining the polarization
vectors $\ep^{(0)}$ and $\ep^{(1)}$. 
The normalizations can be adjusted so that there are no divergences in this limit and
the two polarization vectors become degenerate, both being proportional to the same vector.
There is in fact no state that satisfies $\ep \cdot p \ne0$, a 
curious implication of the existence of a nontrivial Lorentz-violation parameter.
A direct consequence of this is that $\Pi^0 = - \partial \cdot A$ vanishes identically,
thus obstructing
covariant quantization.
This example illustrates how the mass regulator alleviates this degeneracy issue and
ensures that the covariant quantization procedure is consistent.

\section{Summary} 
Direct application of the Gupta-Bleuler covariant quantization procedure to
SME photon sector runs into fundamental issues.
Specifically, the conjugate momentum $\Pi^0$ introduced by adding the gauge-fixing term 
is generally forced to be zero in the presence of
the perturbation term, making the covariant quantization procedure fail.
On the other hand,
covariant quantization of the photon field appears to be consistent when a mass term
is included in the Lagrangian.
This is certainly the case when the
mass term dominates the Lorentz-violating terms $m \gg \sqrt{k_F} p_0$.
In fact, in all of the specific examples that we looked at,
any nonzero mass term actually ensured the existence of
one time-like and three space-like polarization vectors
for arbitrary three-momentum.  
We hypothesize that this is the case for arbitrary perturbative $k_F$ choices.
As future work, it would be of significant interest to confirm or deny this hypothesis.
If true,
then it implies that a consistent method for performing quantum-field-theoretic calculations
in the SME photon sector can be implemented by including a mass regulator term
and taking the appropriate zero mass limit at the end of the calculation.
This procedure is already commonly applied to regulate infrared divergences in QED.

\section{Acknowledgments}
R.~P.\ wishes to thank the kind hospitality of New College of Florida
were part of this work was carried out.
R.~P.\ kindly acknowledges partial financial support by the Portuguese
Funda\c c\~ao para a Ci\^encia e a Tecnologia.

%\appendix*
\section{Appendix: BRST and the condition on the physical states}

The structure of the space of states is most clearly established by
completing the photon lagrangian (\ref{lagrangian-A})
by adding the contributions from a Stueckelberg scalar field
$\phi$ as well as the (anticommuting)
ghost and antighost fields $c$ and $\bar c$
\citep{cambiaso}:
\begin{equation}
\mathcal{L}_{Stueck}=\mathcal{L}_A+\mathcal{L}_\phi+\mathcal{L}_{gh}
\label{lagrangian-Stueck}
\end{equation}
where $\mathcal{L}_A$ is given by (\ref{lagrangian-A}) and
\begin{equation}
\mathcal{L}_\phi=\half(\partial_\mu\phi)^2-\half \xi m^2\phi^2
\end{equation}
and
\begin{equation}
\mathcal{L}_{gh}=-\bar c(\partial^2+\xi m^2)c.
\end{equation}
Lagrangian (\ref{lagrangian-Stueck}) can now be obtained from
the lagrangian
\begin{equation}
\mathcal{L'}_{Stueck}=-\quar F_{\mu\nu} F^{\mu\nu}-\quar k_F^{\mu\nu\alpha\beta}F_{\mu\nu}F_{\alpha\beta}
+\half m^2 (A_\mu- \frac{1}{m}\partial_\mu\phi)^2
+\frac{\xi}{2}B^2+B(\partial_\mu A^\mu+\xi m\phi)
-\bar c(\partial^2+\xi m^2)c
\label{lagrangian-Stueck-2}
\end{equation}
upon integrating out the (auxiliary) Nakanishi-Lautrup field B,
introduced for consistency of the BRST transformation.

Lagrangian (\ref{lagrangian-Stueck-2}) is invariant under the
BRST transformation $s$ defined by
\begin{align}
sA_\mu&=\epsilon\partial_\mu c \label{sA}\\
s\phi&=\epsilon m c\\
s c&=0\\
s \bar c&=\epsilon B\\
s B&=0 \label{sB}
\end{align}
where $\epsilon$ is some infinitesimal grassman-valued gauge parameter.
Note that $s$ is nilpotent off-shell: $s^2=0$.
The general structure is in fact unaltered by the presence of the Lorentz-violating
terms, mainly since gauge invariance is preserved by the $k_F$ term.

As is customary, we now define the set of \textit{exact states} $\mathcal{H}_2$
as the states that can be obtained as the image of $s$ acting on some other state.
If we substitute the field $B$ through its equation of motion by
$-\xi^{-1}(\partial_\mu A^\mu+\xi m\phi)$ we see that  $\mathcal{H}_2$
consists of the states built of the vacuum by acting with the field oscillators
either of the ghost $c$ or of the field combination $\partial_\mu A^\mu+\xi m\phi$.
The set of \textit{physical states} $\mathcal{H}_0$
is defined as the set of (closed) states annihilated by the
BRST operator $s$ modulo $\mathcal{H}_2$.
Note that this procedure makes sense because any exact state,
while closed (because $s^2=0$), is automatically
orthogonal to any closed state: $\langle exact|closed\rangle=0$.
The remaining set $\mathcal{H}_1$ of \textit{unphysical states} are those that
are not closed.
It consists of states built out of the oscillators of the antighost, 
the longitudinal mode of $A_\mu$ ($\lambda=0$) and the scalar $\phi$,
with the exception of the combination $\partial_\mu A^\mu+\xi m\phi$.
Thus we conclude that the exact mode is a linear combination of the
longitudinal mode and the scalar $\phi$,
while the linear combination orthogonal to this is unphysical.
The (remaining) physical states are the three transverse modes of $A_\mu$.

The only mode for which $p^\mu\epsilon_\mu^{(\lambda)}(\vec p)$
does not vanish is the longitudinal mode $\lambda=0$.
This fact follows from examination of the equation of motion in an observer
frame for which $\vec p=0$ which generally exists when a mass term is present, at least
for the case of time-like momenta.
In this frame, the equation of motion for $\ep_\mu$ reduces to 
\beq
\left( p_0^2 \tilde \eta^\mn  - m^2 \eta^\mn \right) \ep_\nu = 0,
\eeq
where $\tilde \eta^{0i} = \tilde \eta^{i0} = 0$ and $\tilde \eta^{00} = 1$.
It is clear by inspection that there is one time-like polarization satisfying $p_0^2 = m^2$
and three space-like polarizations satisfying $p_0^2 = m^2/(1 - 2 k_{\La F}^{0i0i})$, $i = \{1,2, 3\}$
with no sum,
where the spatial orientation of the coordinate system has been rotated to diagonalize
$\tilde \eta^{ij}$ in the boosted frame.
The parameter $k_{\La F}$ represents the boosted value of the corresponding lorentz-violating
parameters in this special frame.
The structure of the solution space is clearly seen in this frame where the time-like polarization
vector points along the momentum and the polarizations corresponding to the perturbed dispersion relations point in  space-like directions and satisfy the transversality condition $\ep \cdot p = 0$.
An observer Lorentz transformation can then be used to boost back to the original frame
maintaining the properties that $\ep_\mu^{(0)}$ is proportional to $p_\mu$ and the other states
are orthogonal to $p_\mu$.

An important caveat to the above argument is that the boost cannot get too large or
the $k_{\La F}$ parameters may grow to order one leading to a breakdown of the perturbative
theory (also called boosting to a nonconcordant frame \cite{lehnert2}).
In the present context, this can ocurr when the photon energy $p_0 \sim m / \sqrt{k_F}$,
where $k_F$ is of order of the elements of the lorentz-violating couplings.

In the massless case,
the longitudinal mode $\lambda=0$ is no longer time-like,
but light-like, as $p^2=m^2 =0$ for this state.
Any exact mode ($\partial_\mu A^\mu \ne0$) will then no longer involve
the longitudinal mode, but possibly one of the other modes,
which must necessarily satisfy light-like dispersion relation $p^2=0$,
due to (\ref{transversality})
(because otherwise we would have $\epsilon\cdot p=0$).
 In fact, generically this mode does not even exist as is shown in the special example
 presented in the body of the paper. 
The  problem arises when the equation of motion is applied with $p^2=0$ resulting in the
condition $K^{\mn} \ep_\nu = 0$.
It is easy to demonstrate that $K^\mn$ is generically of rank 3 (except for certain special
choices of $\vec p$) which implies that the
only polarization with $p^2=0$ is the gauge state with 
polarization proportional to the momentum.
This leads to the conclusion that there is no exact state in these cases.

\bibliography{photonquant_RP2}

\end{document}